\documentclass[%
 reprint,
 superscriptaddress,
 amsmath,amssymb,
 aps,
]{revtex4-2}

\usepackage{graphicx}
\usepackage{dcolumn}
\usepackage{bm}

\begin{document}


\title{Topological bands in metals} 
\author{Yu. B. Kudasov}
\email{yu\_kudasov@yahoo.com}
\affiliation{Sarov Physics and Technology Institute NRNU "MEPhI", 6, str. Dukhov, Sarov, 607186, Russia}


\begin{abstract}

In crystalline systems with a superstructure, the electron dispersion can form a nontrivial covering of the Brillouin zone. It is proved that the number of sheets in this covering and its monodromy are topological invariants under ambient isotopy. As a concrete manifestation of this nontrivial topology, we analyze three-sublattice models for 120$^\circ$-ordered helimagnets in one, two, and three dimensions. The two-dimensional system exhibits unconventional $f$-wave magnetism and a specific topological metal state characterized by a spin-textured, one-sheeted Fermi surface. The observable transport signatures of the topological metal and its potential experimental realization are briefly discussed.
\end{abstract}

\maketitle

The topology of electronic band structures has been a central topic in solid-state physics in recent decades.\cite{Bansil,Hasan} Research in this area has largely focused on topological insulators and their edge states,\cite{Hasan,Tkachev} as well as on systems with topological defects, such as Dirac and Weyl semimetals.\cite{Armitage} It has been shown that nontrivial topological properties give rise to observable transport phenomena, including highly mobile surface charge carriers, anomalous and nonlinear Hall effects.\cite{Du,Nandy} The Berry (geometric) phase \cite{Berry} and derived quantities, such as the Berry vector potential and curvature, are key theoretical tools for describing these phenomena. Combined with symmetry analysis, they enable the derivation of topological invariants and the comprehensive classification of topological phases.\cite{Chiu}

Magnetic topological insulators and semimetals, including helimagnets, have also been studied.\cite{Bernevig,Yao} At the same time, specific nontrivial band structures in metallic helimagnets have been identified that are distinct from these established classes. In this Letter, we demonstrate that they are unrelated to the Berry phase or topological defects, and we illustrate this finding with simple examples of helimagnetic systems.   

The electronic band structure of metallic helimagnets has been studied extensively.\cite{Dzyaloshinskii,Brinkman, Fresard,Fraerman,Kishine,Kudasov} Both approximate methods \cite{Dzyaloshinskii,Calvo2,Kudasov} and exact solution \cite{Calvo} have established that an effective helical magnetic field produces a characteristic gapless multiband structure. 

Helical magnetic systems typically comprise two periodic structures. Let $\mathbf{t}$ be a primitive vector of the crystal lattice such that a translation by this vector rotates the magnetic (spin) system by an angle of $\alpha_m = 2 \pi/m$ ($m>1$). Here and below, we discuss only commensurate magnetic structures, i.e., $m \in \mathbb{N}$. The primitive translation vectors of the system as a whole are then determined by the magnetic superstructure (e.g., $\mathbf{T} = m \mathbf{t}$). Bloch's theorem allows for the construction of a magnetic Brillouin zone and yields periodic dispersion relations in reciprocal space.

However, within the framework of spin space group theory,\cite{Brinkman} a translation by vector $\mathbf{t}$ combined with a rotation of the magnetic system by an angle $\alpha_m$, being a symmetry operator ($\hat{\mathbf{t}} \hat{\mathbf{r}}_m$), leads to a generalized Bloch theorem \cite{Sandratskii} and an extended Brillouin zone. As shown below, the presence of the two commensurate periodic structures has profound consequences for the topology of the band structure.
  
It has been proved recently that the electron dispersion in commensurate helimagnets has a symmetry related to time reversal:\cite{KudasovPRB}
\begin{eqnarray}
	\varepsilon _ {\mathbf{k}, \langle \boldsymbol\sigma \rangle} = \varepsilon _ {-\mathbf{k},- \langle \boldsymbol\sigma \rangle} \label{disp}
\end{eqnarray} 
where $\mathbf{k}$ is the wave vector and $\langle \boldsymbol\sigma \rangle$ is the expectation value of the spin projection. 
A Kramers-like degeneracy exists throughout the Brillouin zone, except at special points defined by the following conditions: $\exp\big(i \mathbf{k} \mathbf{T}\big)=-1$ for even $m$ and $\exp\big(2 i \mathbf{k} \mathbf{T}\big)=1$ for odd $m$.\cite{KudasovPRB} 

The realization that spin-orbit-free compensated magnets with strong band spin splitting -- such as altermagnets in collinear structures \cite{Yuan,Smejkal} and unconventional magnets in noncollinear ones \cite{Brekke,Ezawa} -- constitute a new, promising class of functional materials has renewed interest in these systems.\cite{Song,Sears} The helimagnets with even $m$ possess the symmetry operator $\hat{\mathbf{T}}_{1/2}\hat{\mathbf{\theta}}$, where $\hat{\mathbf{T}}_{1/2}$ is the translation by $\mathbf{T}/2 =m\mathbf{t}/2$ and $\hat{\mathbf{\theta}}$ is the time-reversal operator.\cite{KudasovPRB} This symmetry gives rise to unconventional magnetism (e.g., of a $p$-wave type \cite{Brekke}).

Let us consider the states of a single electron in a crystal lattice, i.e., in a periodic potential $U(\mathbf{r}+\mathbf{t}) = U(\mathbf{r})$, where $\mathbf{t}$ is a  primitive vector of the Bravais lattice.   The potential $U(\mathbf{r})$ is assumed to be a scalar (real) or spinor function. The Schr\"{o}dinger equation in this case has the form  
\begin{eqnarray}
\hat{H} \varphi(\mathbf{r})=\left[-\frac{\Delta}{2} + U(\mathbf{r})\right] \varphi(\mathbf{r})= \varepsilon \varphi(\mathbf{r}). \label{H}
\end{eqnarray}
Eigenvalues $\varepsilon_{i}(\mathbf{k})$ and eigenvectors $\varphi_{i}(\mathbf{k})$ of the Hamiltonian are defined by the wave vectors $\mathbf{k}$. In the limit of an infinite crystal, these functions are continuous and, according to Bloch's theorem, periodic in $k$-space:\cite{Ashcroft}
\begin{eqnarray}
	\varepsilon_{i}(\mathbf{k}+\mathbf{K})=\varepsilon_{i}(\mathbf{k}), \label{p1}\\
	\varphi_{i}(\mathbf{k}+\mathbf{K})= \varphi_{i}(\mathbf{k}) 
\label{p2}
\end{eqnarray}
where $\mathbf{K}$ is a vectors of the reciprocal lattice. Here, the index $i$ enumerates the solutions of Eq.~(\ref{H}). 

Taking into account the boundary conditions Eqs.~(\ref{p1}) and (\ref{p2}), the Brillouin zone can be represented as a closed smooth manifold $B$, which is topologically equivalent to a torus $T^n$ with $n=1,2,3$ for 1D, 2D, and 3D systems, respectively.

The Bloch wave function can be expanded as follows: $\varphi_{i}(\mathbf{k}) = \sum_{j=1}^{n} c_{ij}(\mathbf{k}) \psi_{ij}$ where $\psi_{ij}$ form an orthonormal basis in the coordinate space, $c_{ij}(\mathbf{k})$ are the smooth complex functions of $\mathbf{k}$,  and $n>1$. Assuming a finite number of the terms in the sum and taking into account the normalization condition $\sum_{j=1}^{n} |c_{ij}|^2=1$, one can associate  $\varphi_{i}(\mathbf{k})$ at fixed $\mathbf{k}$ and $i$ with a point on a $(2n-1)$-sphere ($S^{2n-1}$). In the case of an infinite number of the terms, $S^{2n-1}$ is replaced by the infinite-dimensional Hilbert space $\ell^2$. Consequently, $\varphi_{i}(\mathbf{k})$ defines a closed smooth manifold $\varphi$ in a space $X \cong B \times G$, where $B \cong T^n$ and $G$ is topologically equivalent to $S^{2n-1}$ or $\ell^2$. $\varphi$ and $B$ are obviously of the same dimension.

 The eigenvectors $\varphi_{i}(\mathbf{k})$ are mutually orthogonal. Therefore, for any given $\mathbf{k}$, they correspond to distinct points in the fiber $G$, ensuring that the sheets $\varphi_{i}$ and $\varphi_{j}$ ($i \ne j$) do not intersect. This allows us to assume that the map
$f: \varphi \rightarrow B$ is regular, i.e., has a nowhere vanishing Jacobian:\cite{note1}
\begin{eqnarray}
	\det\bigg ( \frac{\partial k_{\alpha}}{\partial x_{\beta}}\bigg) \ne 0  , \label{Jacobian}
\end{eqnarray}    
where $k_{\alpha}$ and $x_{\beta}$ are the local coordinates defined in $B$ (the Brillouin zone) and $\varphi$, respectively.
    
The energy dispersion $\varepsilon_{i}(\mathbf{k})$ can also be represented as a manifold $\varepsilon$ in the space $B \times I$, where $I$ is the unit interval  $[0,1]$ (under the assumption that $\varepsilon_{i}(\mathbf{k})$ bounded). However, the map
$f: \varepsilon \rightarrow B$ can be singular because the eigenvalues can be degenerate. 

\begin{figure}
	\includegraphics[scale=0.9]{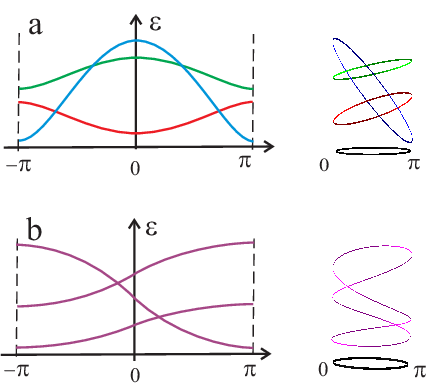}
	\caption{\label{f1} 1D band structures and their topological classification. The right panels provide a schematic topological representation, depicting the Brillouin zone as a circle ($S^1$).}
\end{figure}

Under the conditions stated above, the map $f: \varphi \rightarrow B$ is a covering map. In solid-state physics, usually only trivial coverings are considered; that is, $\varphi \cong B \times F$ where $F$  is a countable or finite discrete topological space. In this case, each sheet $\varepsilon_{i}(\mathbf{k})$ is a closed manifold topologically equivalent to the base $B$. Consequently, the index $i$ in Eqs.~(\ref{p1}) and (\ref{p2}) specifies the indecomposable coverings, as shown in Fig.~\ref{f1}a.  

The complete inverse image $f^{-1}(k)$ (the fiber of the covering) of any point $k \in B$ consists of $m$ points. This number is referred to as the number of sheets. An indecomposable covering with $m>1$ is nontrivial. Since the map $f$ is regular, the number of sheets for an indecomposable covering must be finite.\cite{Dubrovin}  

Let $\gamma$ be a loop in $B$ starting and ending at $k_0$, i.e., $\mathbf{k}_0 \rightarrow \mathbf{k}_0 +\mathbf{K}$ in terms of $\mathbf{k}$-space, and let $\{x_1, ..., x_m\}$ be the fiber over $k_0$ for an indecomposable covering. For each starting point $x_i$ in this fiber, there is a unique covering path $\mu(x_i)$ in $\varphi$ that covers $\gamma$. The endpoint of $\mu(x_i)$ is some $x_j$ in the same fiber; however, $\mu(x_i)$ is not necessarily a loop (if $i \ne j$). A schematic 1D example of the dispersion curve corresponding to nontrivial $\varphi$ is shown in Fig.~\ref{f1}b.

\begin{figure*}
	\includegraphics[scale=0.38]{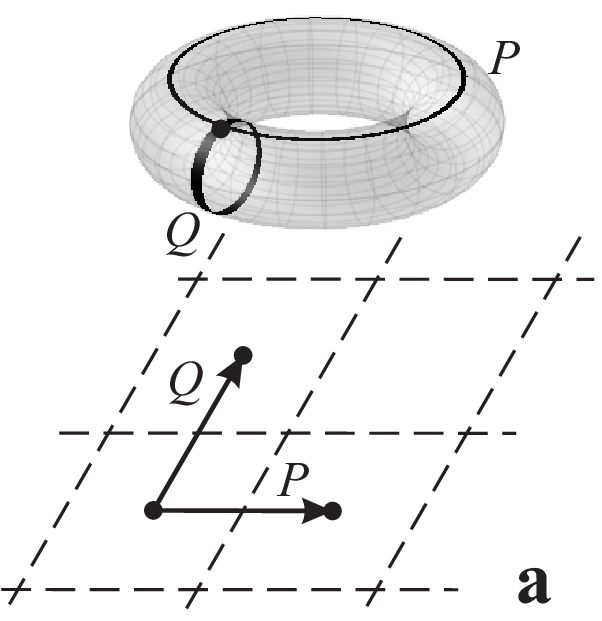} \; \; \; \; \; \;
	\includegraphics[scale=0.3]{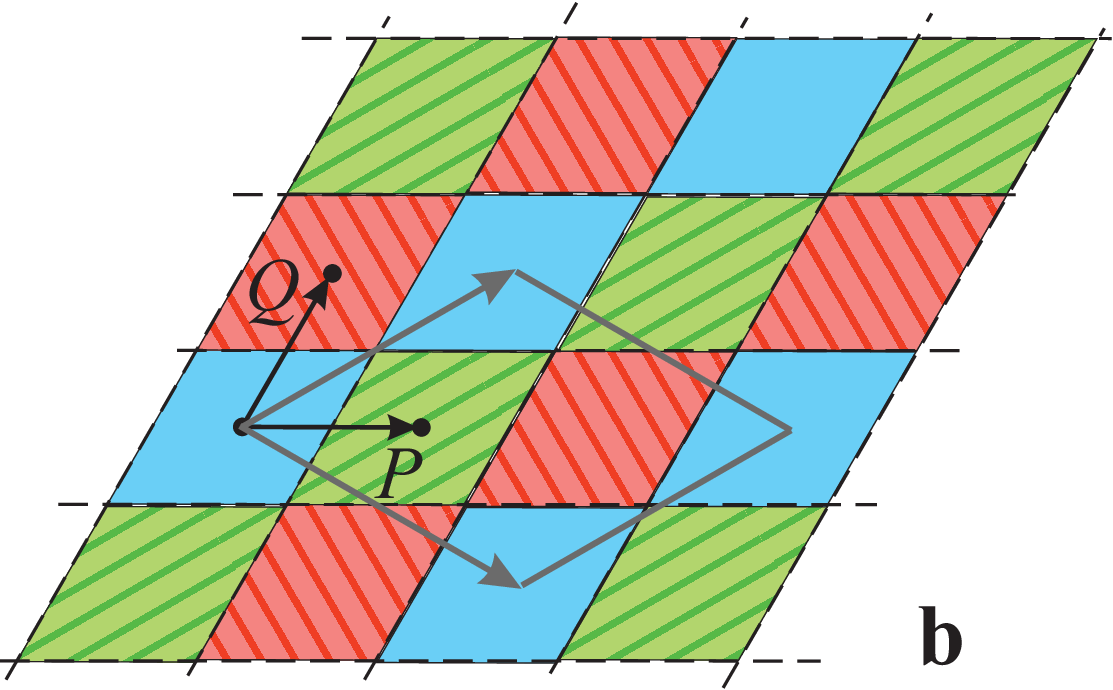} \; \; \; \; \; \;
	\includegraphics[scale=0.25]{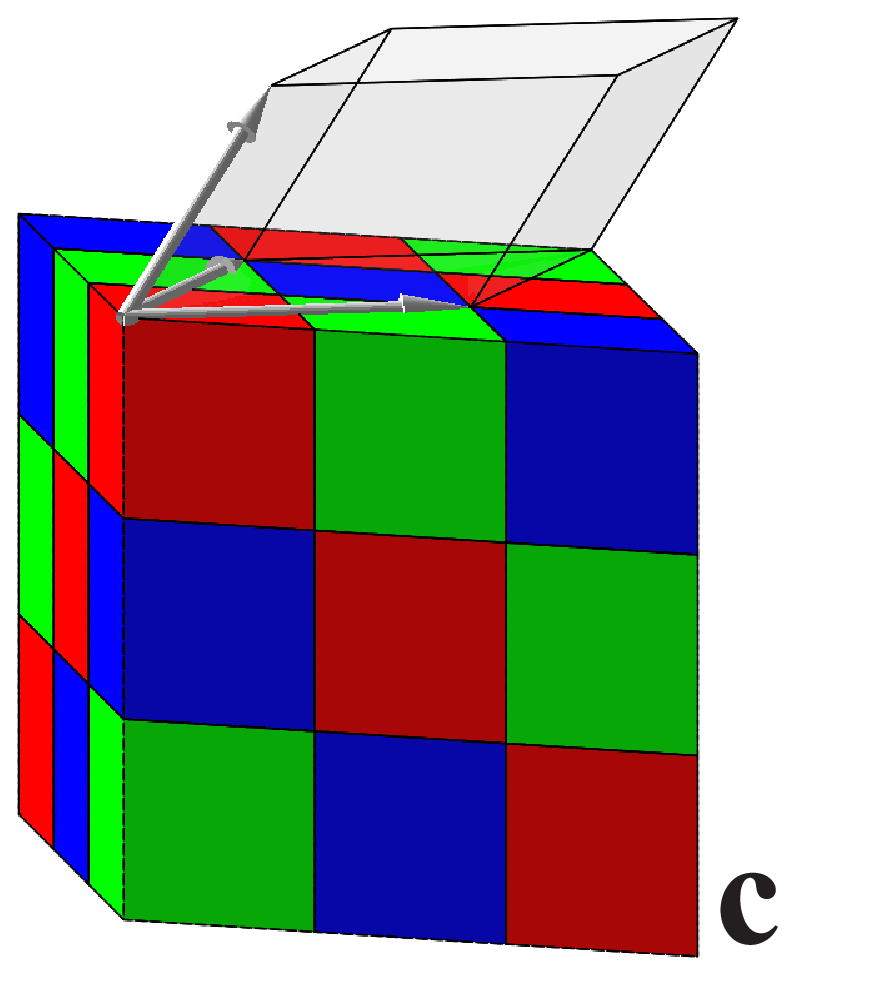}
	\caption{\label{f2} Schematic views of 2D and 3D coverings: (a) plane representation of torus (base), (b) plane representation of a 3-sheeted covering over $T^2$, and (c) solid representation of a 3-sheeted covering over $T^3$. The sheets of covering spaces are distinguished by color and texture.}
\end{figure*}

Since any loop within the Brillouin zone $B$ is homotopic to a null path (i.e., contractible to a point), the choice of the starting point is inessential. For this reason, we omit the starting point in the notation, where possible.

$\mathbb{R}^2$ is the universal covering space of the torus, therefore it is convenient to use a plane representation for 2D systems, as shown in Fig.2a. The horizontal and inclined lines in this figure schematically denote the boundaries of the Brillouin zone. Gluing them together in the usual way forms a torus.\cite{Kalajdzievski} The path from the center of any cell to that of an adjacent one is a non-contractible loop on the torus. A pair of such paths, which are generators of the fundamental group of the torus, are shown ($P$ and $Q$) in Fig.2a. As an example, consider the three-sheeted covering of the torus, which corresponds to the model discussed below. The different sheets are indicated by colors. It is clear that the lifts of paths $P$ and $Q$ are not closed, because there is a transition to another sheet. It should be emphasized that the dashed lines and coloring in Fig.\ref{f2}a,b are guides for the eye; in effect, there are no singular points or lines. The plane representation of the covering is also periodic. The paths $PQ$ and $P^2 Q^{-1}$ are generators of the fundamental group of the covering space.\cite{KudasovJETPL2} This group is isomorphic to the fundamental group of the torus, i.e., $\pi_1(\varphi)\approx \pi_1(B)$.

$\mathbb{R}^3$ is the universal covering space of $T^3$. Let the 3D Brillouin zone be a parallelepiped. A schematic representation of a three-sheeted covering over this base is shown in Fig.~\ref{f2}c, where coloring again denotes the different sheets of the covering space. The edges of the unit parallelepiped correspond to the generators of $\pi_1(B)$, and the grey arrows indicate the generators of $\pi_1(\varphi)$.

The curve in the right panel of Fig.~\ref{f1}b resembles a knot, particularly considering that $\varphi$ has no self-intersections. It is well-known that embedding of a circle into the $\mathbb{R}^3$ as a knot leads to the question its nontriviality.\cite{Hatcher} 
Therefore, the notion of topological equivalence for the coverings discussed above requires precise clarification.

\textbf{Proposition}. Let $\varphi$ be a subspace of $X \cong B \times G$, where $B$ and $G$ are path-connected topological spaces, and let the restriction of the projection $\pi: X \rightarrow B$ be a regular covering map $p: \varphi \rightarrow B$. Then, if $\tilde{\varphi}$ is ambient isotopic to $\varphi$ within $X$,  $\tilde{\varphi}$ is also a covering space (under the restriction of $\pi$) which is equivalent to $\varphi$.

\textit{Proof}. The ambient isotopy is a continuous mapping $F: X \times I \rightarrow X$ such that $F_t$ is a homeomorphism for every $t \in I$ with $F_0 = \text{id}_X$, and $F_1(\varphi) = \tilde{\varphi} $.\cite{Kalajdzievski} 
Define the map $\tilde{p}:\tilde{\varphi} \rightarrow B$ as the composition
\begin{eqnarray}
	\tilde{p}= p \circ \bar{F}_1 |_{\tilde{\varphi}}, \label{monodromy}
\end{eqnarray}   
where $\bar{F}$ is the inverse ambient isotopy, i.e., $\bar{F}_t F_t = \text{id}_X$. A composition of a homeomorphism and a covering map is again a covering map. Therefore, $\tilde{p}$ is a covering map. Furthermore, $\tilde{p}$ and $p$ are equivalent coverings \cite{Armstrong} by definition Eq.~(\ref{monodromy}).$\blacksquare$  

\textit{Corollary 1}. Monodromy representations of $p$ and $\tilde{p}$ are isomorphic.

\textit{Corollary 2}. Since the covering map $p$ induces a monomorphism $p_*: \pi_1(\varphi,x_0) \rightarrow \pi_1(B,k_0)$, the indecomposable covering can be nontrivial only if the fundamental group of the base $\pi_1(B, k_0)$ is nontrivial. This condition is met in case of the Brillouin zone: $\pi_1(T^n) \approx \mathbb{Z}^n$. Thus, the topology of the base space, e.g. $B = T^n$, is the source of the nontriviality of the covering map.

Let us translate the above results into band structure theory. If $\varphi(\mathbf{k})$ is a $m$-sheeted regular covering over the Brillouin zone, an individual dispersion sheet is non-periodic within the Brillouin zone, and the conditions Eqs.~(\ref{p1}) and (\ref{p2}) are satisfied due to permutations of the dispersion sheets, that is, by monodromy, as schematically shown in Fig.~\ref{f1}b. The number of sheets $m$, as well as the sequence of the sheet permutations are topological invariants. Since $m$ is finite, a superstructure must exist in the system, i.e., the covering space itself forms a periodic structure in the $\mathbf{k}$-space with a period commensurate with that of the Brillouin zone. 

\begin{figure}
	\includegraphics[scale=0.44]{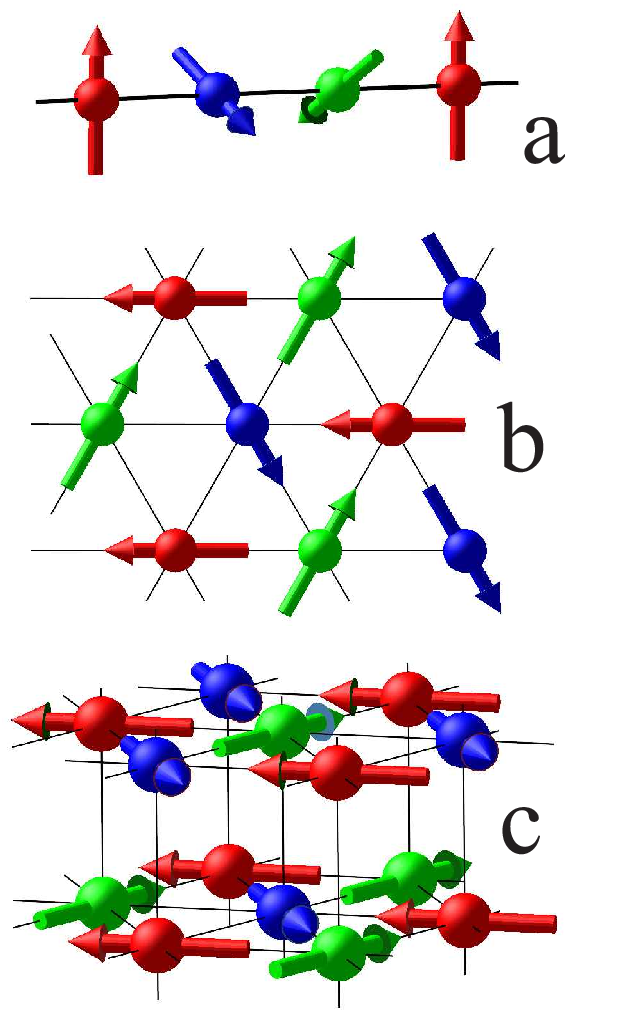}
	\caption{\label{f3} Helical structures with the 120$^\circ$-order on (a) 1D, (b) 2D, and (c) 3D lattices. The magnetic field at the sites is indicated by color and arrows.}
\end{figure}

The nontrivial band structure discussed above can be demonstrated with a tight-binding model of a helimagnetic metal. The Hamiltonian has the same general form for 1D, 2D, and 3D lattices:

\begin{eqnarray}
	\hat{H}_{3sl}=	-\sum_{\langle i,j\rangle, \sigma}\left( \hat{a}^\dagger_{i\sigma} \hat{a}_{j\sigma} + \text{h.c.}\right) \nonumber\\ -
\mathbf{h}_{i} \sum_{i,\sigma,\sigma^\prime} \hat{a}^\dagger_{i\sigma} \hat{\boldsymbol{\sigma}} \hat{a}_{i
	\sigma^\prime}  \label{H3sl}
\end{eqnarray}
where $\hat{a}^\dagger_{i\sigma} (\hat{a}_{i\sigma} )$ is the creation (annihilation) operator for an electron with spin projection $\sigma = \uparrow, \downarrow$ at the i-th site, $\hat{\boldsymbol{\sigma}}$ are the Pauli matrices, and $\mathbf{h}_{i}$ is the (effective) magnetic field at the i-th site. The notation $\langle \ldots \rangle$ denotes the sum over nearest-neighbor pairs. We consider a three-sublattice model with a 120$^\circ$ magnetic order. In this configuration, all $\mathbf{h}_{i}$ vectors are coplanar with constant magnitude $|\mathbf{h}_{i}|=h_0$ and the angle between the vectors on different sublattices is $\pm 2 \pi/3$. Furthermore, all nearest-neighbor sites belong to different sublattices. Since spin-orbit coupling is absent in the model, the magnetic plane, in which the effective field lies, can be chosen arbitrarily. The 120$^\circ$ magnetic order is schematically shown in Fig.~\ref{f3} for 1D chain, 2D hexagonal, and 3D simple hexagonal lattices. A primitive crystallographic translation rotates the entire magnetic structure by $\pm 2 \pi/3$.  An explicit matrix form of the Hamiltonian Eq.~(\ref{H3sl}) is presented in the Supplemental material.

\begin{figure}
	\includegraphics[scale=0.35]{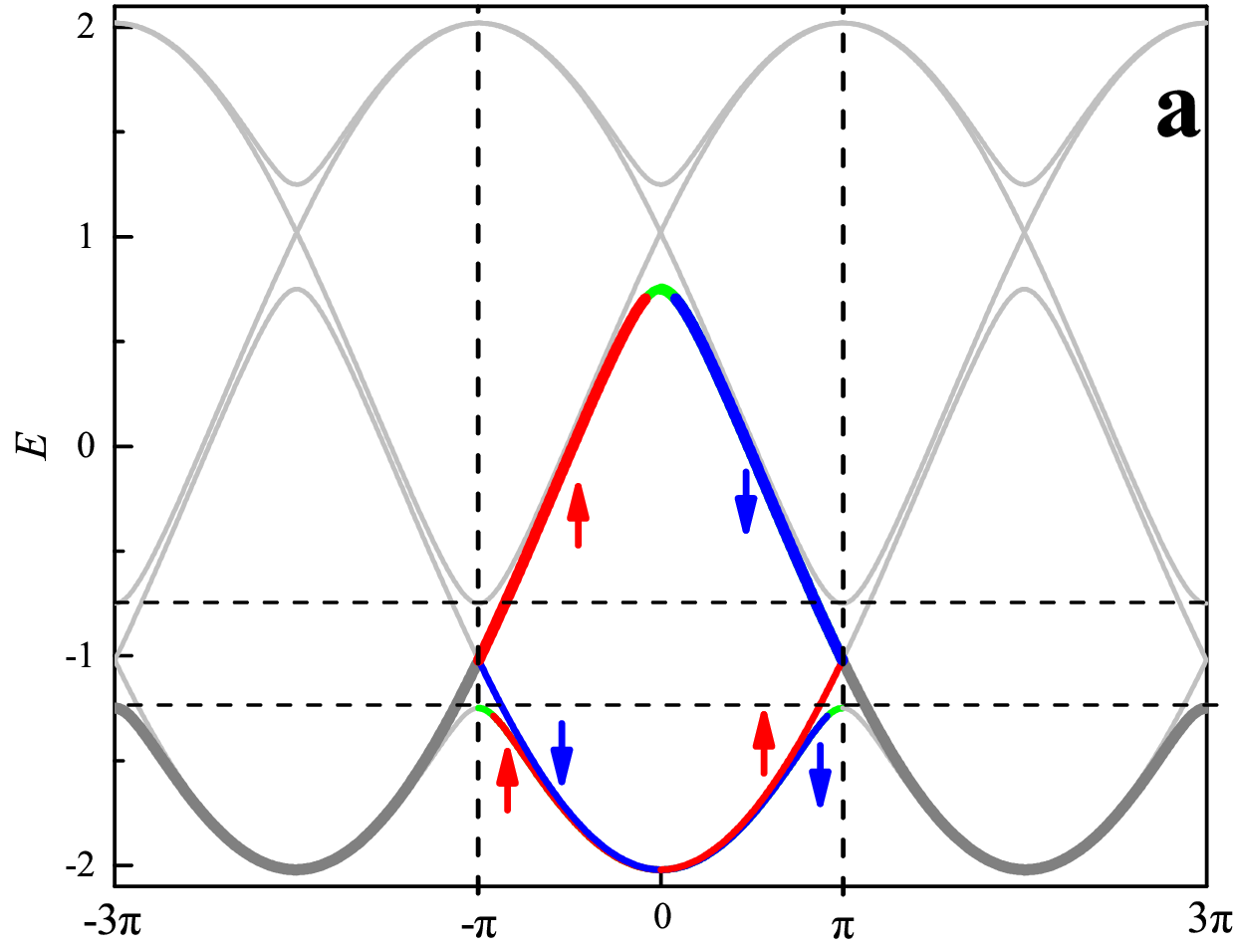}  \includegraphics[scale=0.35]{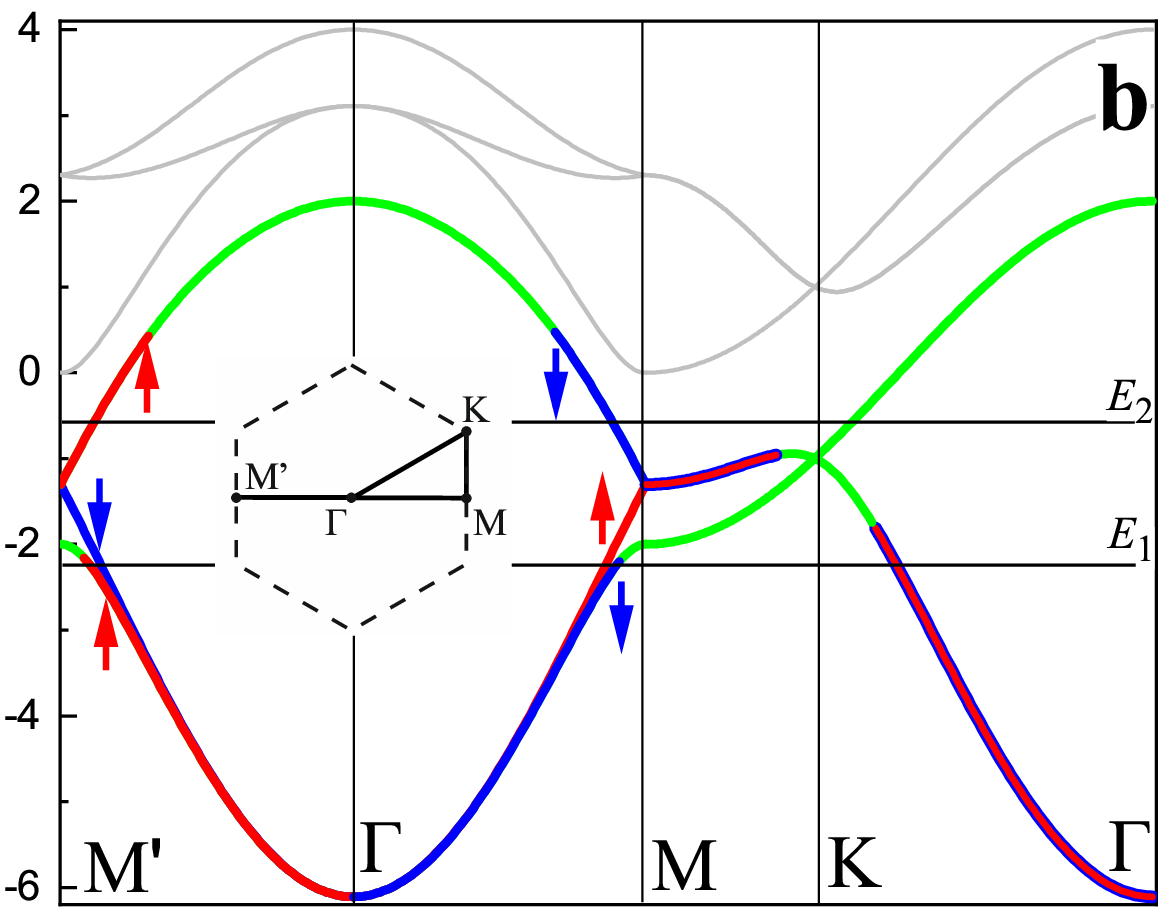}
	\caption{\label{f4} The band structure in the tight-binding model for (a) 1D chain ($h_0=0.25$) and (b) 2D hexagonal ($h_0=1$) lattices. The average spin projection on the axis perpendicular to the magnetic plane is indicated by color and arrows: red and blue if $|\langle \hat{\mathbf{\sigma}}_z\rangle|>1/2$, green otherwise.}
\end{figure}

The dispersion curves obtained from the 1D tight-binding model are shown in Fig.~\ref{f4}a within the ordinary (magnetic) ($k \in [-\pi,\pi]$), and extended (crystallographic) Brillouin zones ($k \in [-3\pi,3\pi]$). The lower band marked by the thick line unambiguously corresponds to a continuous $\varphi_{i}(\mathbf{k})$. It is periodic within the extended Brillouin zone. Band folding results in three non-periodic dispersion sheets within the magnetic (ordinary) one. The magnetic structure is shown by color and arrows within the magnetic Brillouin zone. It should be noted that the commensurate helical magnetic field lifts the spin degeneracy of the band structure but conserves the Kramers-like symmetry Eq.~(\ref{disp}). 

If the Fermi level falls into the energy range between the horizontal dashed lines in Fig.~\ref{f4}a, a special state emerges: a single non-periodic band crosses the Fermi level. We refer to this as a topological metal because such a structure is a consequence of a nontrivial covering over the Brillouin zone. One can see that a backward scattering without spin-flip is forbidden and that a persistent spin current exists in this case.\cite{Kudasov}

\begin{figure}
	\includegraphics[scale=0.3]{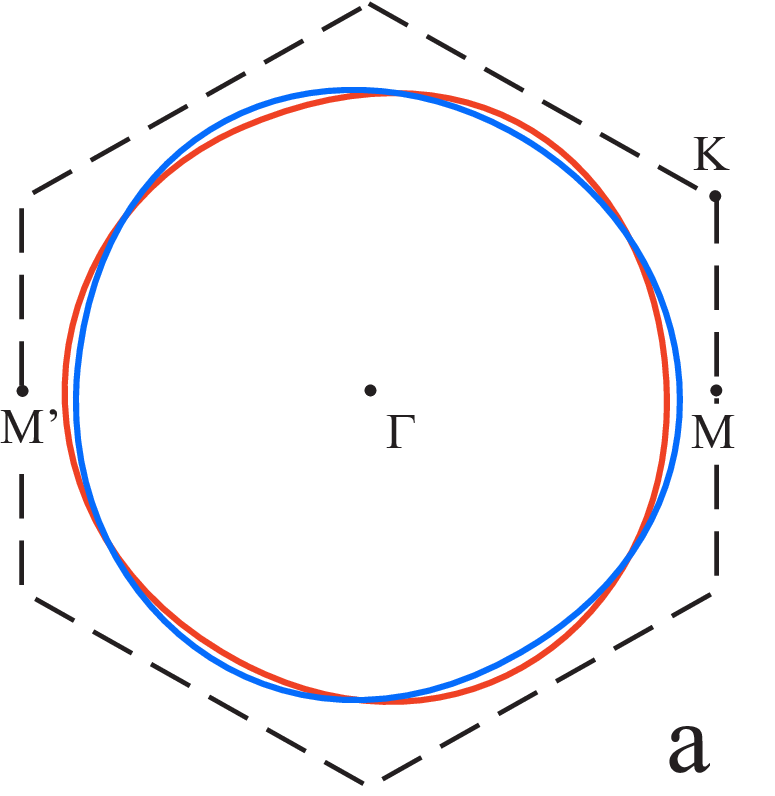} \; \; \;	\includegraphics[scale=0.3]{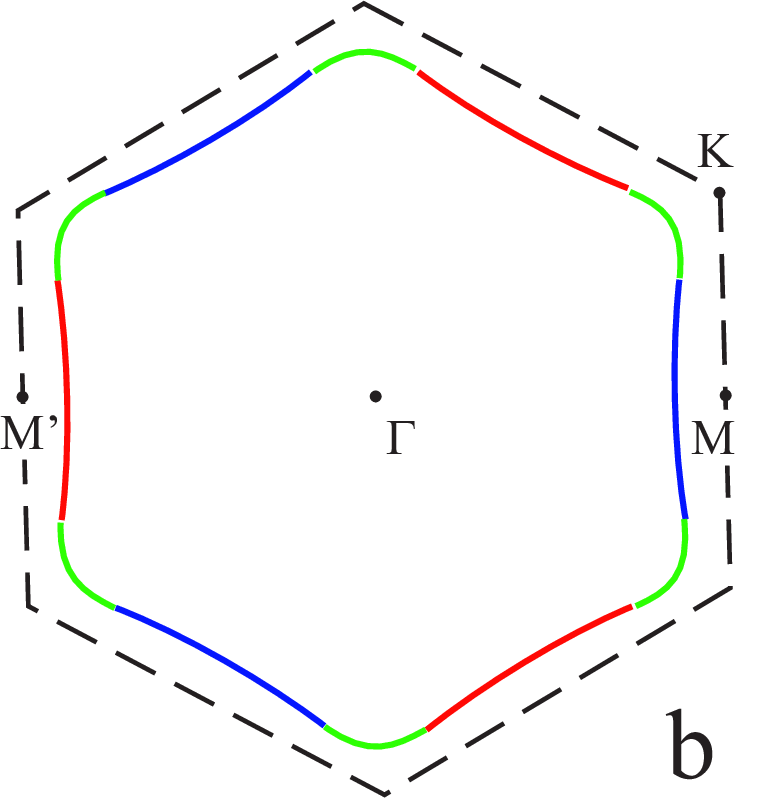}
	\caption{\label{f5} Fermi surface in the 2D tight-binding model for Fermi levels corresponding to (a) $E_1$ and (b) $E_2$ in Fig.~\ref{f4}b. The average spin projection is indicated in the same manner as in Fig.~\ref{f4}.}
\end{figure}

The band structure of the tight-binding 2D model is shown in Fig.~\ref{f4}b. The lower bands also exhibit a 3-sheeted structure. Along the M'-$\Gamma$-M direction, the lower dispersion curves are similar to those of the 1D model (Fig.~\ref{f4}a). The shapes of the Fermi surface for the Fermi level at $E_1$ and $E_2$ are presented in panels (a) and (b) of Fig.~\ref{f5}, respectively. The spin splitting of the Fermi surface in Fig.~\ref{f5}a leads to the 2D $f$-wave magnet.\cite{Ezawa}

If the Fermi level lies at $E_2$  (Fig.~\ref{f5}b), a single spin-textured Fermi surface appears, with the texture obeying Eq.~(\ref{disp}). This contrasts with altermagnets and unconventional magnets (e.g. of $p$- and $f$-wave type) with the trivial topology, where the spin splitting leads to pairs of bands and a Fermi surface with an even number of sheets.\cite{Smejkal,Ezawa} We refer to this state as a 2D topological metal. It demonstrates unusual transport behavior: the suppression of both backscattering without spin-flip and umklapp electron-phonon scattering leads to a drastic increase in conductivity.\cite{Kudasov}

The 3D tight-binding model also features nontrivial bands. The band structure along the $\mathbf{a^*}$ direction corresponding to M'-$\Gamma$-M in the 2D model \cite{Hart} is similar to Fig.~\ref{f4}b (see Supplemental Material). However, a detailed discussion of the 3D model lies beyond the scope of the present work.

To practically implement the concept of a topological metal in a real material, several conditions must be met. For instance, in a layered crystal, highly conductive layers must alternate with magnetic layers possessing 120$^\circ$ ordering to induce the effective magnetic field. Furthermore, the Fermi level must be located in an energy region with a single non-periodic band. PdCrO$_2$, a metallic delafossite with a complex magnetic order and anomalously high conductivity,\cite{Mackenzie} has been considered a candidate for the topological metal. Non-reciprocal transport \cite{Akaike} and an unusual anomalous Hall effect \cite{Takatsu} have also been observed in this compound. Evidence of the topological metal state in this substance could be provided by observing the spin texture of the reconstructed Fermi surface, for example using spin-resolved angle-resolved photoemission spectroscopy.\cite{Dil} Another way to realize the topological metal is by creating an artificial structure with alternating magnetic and conductive layers, for example, in van der Waals systems.\cite{Chang}

While a nontrivial band structure can also exist in non-magnetic systems (e.g., in crystals with helical symmetry \cite{Damnjanovic}), the topological metal state appears only in helimagnets due to the combination of lifted spin degeneracy and the Kramers-like symmetry Eq.~(\ref{disp}).

\section*{Supplementary Material}
An explicit form of the tight-binding models used in the text and the band structure for the 3D model are provided. 

\section*{Acknowledgments}

This work was supported by the National Center for Physics and Mathematics (Project No. 7 “Investigations in high and ultrahigh magnetic fields").

\nocite{*}
\bibliography{disp_top_bib1} 

\begin{thebibliography}{40}
\bibitem [{not()}]{note1}%
  \BibitemOpen
  \href@noop {} {}\bibinfo {note} {This is almost always true (apart from
  special cases, e.g. topological defects in the Brillouin zone like in Dirac
  and Weyl systems).}\BibitemShut {Stop}%
\end{thebibliography}

\end{document}